# A Carbon Nanotube Immunosensor for *Salmonella*

*Mitchell Lerner[1], Brett Goldsmith[1], Ronald McMillon[2], Jennifer Dailey[1], Shreekumar Pillai[2], Shree R. Singh[2], A.T. Charlie Johnson[1]*

[1] University of Pennsylvania, Department of Physics and Astronomy

[2] Center for NanoBiotechnology Research, Alabama State University

Microbial pathogens cause an estimated annual 6.5-81 million cases of food borne human illness in the US, with significant mortality, and costs ranging from $ 2.9 - $ 6.7 billion annually[1]. *Salmonella* is the leading cause of food borne illness in the United States, found on fresh produce in addition to animal hosts such as chicken, pork and seafood[2]. The efficacy of techniques used to detect food borne pathogens including *Salmonella* is limited by the time and sensitivity of the methods used. Newer methods based on PCR are very sensitive but require pure samples, lengthy processing times and expertise in molecular biology[3]. Conventional bacteriological culturing methods to detect *Salmonella* require at least 72 hours for positive identification[4]. Detection methods based on ELISA or PCR are much quicker, but may require pure samples, lengthy processing times, expertise in molecular biology and they could be quite expensive[5]. Rapid and easy detection of food-borne bacteria such as *Salmonella* would be a useful tool for protecting public health. An ideal sensor technology would be small and cheap enough to enable automated detection of foodborne disease at multiple locations along the product chain, from harvest and transportation to point-of-use in the kitchen. Biosensors that combine biological elements for target recognition with an all-electronic, nano-enabled readout element are promising candidates for use as very large sensor arrays that could assist medical diagnosis or other analysis through the simultaneous quantification of hundreds of biological and/or biomolecular targets using a single small-volume sample[6-8]. Therefore, we have begun to investigate a carbon nanotube-based immunosensor to detect *Salmonella*.

There is an extensive literature on carbon nanotube (CNT) based sensors, including applications in medicine and environmental monitoring[8-13]. Advantages of sensor based on carbon nanotube field effect transistors (FETs) include mature electronics development, very high intrinsic sensitivity, and well-understood carbon surface chemistry, which allows flexible functionalization strategies [14-17]. Carbon nanotube transistor devices have been used as all-electronic transduction elements in both highly specific sensors based on antibody-antigen type relationships[7, 8, 18, 19] and in more broadly tuned "nose-like" sensors[11]. It has been reported that large-area (500 µm x 500 µm channel area) transistors based on networks of mixed semiconducting and metallic nanotubes, coated with non-specifically adsorbed anti-*Salmonella* antibodies, can act as detectors for *Salmonella* suspended in a sterile, low-salt solution[7].

In this study, we focused on the use of small-area (2.5 µm x 15 µm channel area) nanotube transistors, since such devices are compatible with the long-term goal of a large array of diverse biosensors integrated in a lab-on-a-chip system. We used a method to covalently bind anti-

*Salmonella* antibodies to the CNT sidewall, since this affords additional control over the conformation and location of the bio-recognition element with respect to the nanotube device. We investigated detection of *Salmonella* in nutrient broth, a sample that more closely resembles the application of detection of foodborne pathogens at harvest or point-of-use. Our findings support the promise of this system for use in real-world detection of *Salmonella*. We find that the ON-state current of the functionalized NT sensor is sharply decreased (to saturation) by exposure to *Salmonella* at concentrations as low as 1000 colony forming units (cfu) per ml of broth, and that this change is specific to *Salmonella*, both results consistent with earlier work using bacteria re-suspended in low-salt diluted buffer [7]. We also observe that the carrier mobility of the functionalized CNT FET shows a marked dependence on bacteria concentration to levels as high as $10^8$ cfu/ml, and that this is *not* specific to *Salmonella*. This study thus represents a significant step towards ultimate use of CNT devices as detectors of foodborne pathogens.

**Methods**

Fabrication of carbon nanotube field effect transistors (CNTFETs) follows previously published standard practices[11, 20]. Briefly, CNTs are grown via chemical vapor deposition (CVD) at 900 °C using a randomly dispersed iron nitrate catalyst. The growth substrate is a p++ doped silicon wafer with a 300 nm thermal oxide. Photolithography is used to define Au/Cr source and drain electrodes on top of the carbon nanotubes [21]. The doped wafer serves as a global backgate electrode for the CNTFETs. The devices have a 2.5 μm source-drain separation.

*Salmonella enterica* serovar Typhimurium (ATCC 13311) was obtained from American Type Culture Collection (ATCC, Manassas, VA), while *Escherichia coli* (85W 0400), *Streptococcus pyogenes* (85W 1180) and *Stpahylococcus aureus* (85W 1178) were obtained from WARD's Natural Science (Rochester, NY). The bacterial cultures were maintained on nutrient agar slants. Bacterial colony forming units (CFU /ml) were determined by the standard plate count as described previously [22]. Briefly, 10 μL aliquots of each bacterial culture were inoculated individually into 5 mL of fresh nutrient broth and incubated overnight at 37 $^0$C while shaking. Serial, 10-fold dilutions of each culture were then created using nutrient broth. 50 μL samples of each dilution were inoculated onto nutrient agar plates, and incubated overnight at 37 $^0$C. The number of colony forming units was determined by counting colonies manually and applying the appropriate dilution factors. Tubes containing each dilution were then sealed and stored at $4^0$C until testing with CNTFETs.

Polyclonal anti-*Salmonella* antibodies (catalog # 20-SR11; stock concentration of 5 mg/mL) were purchased from Fitzgerald Industries International (Acton, MA). These antibodies are specific to a broad range of *Salmonella* O and H antigens. This solution was then diluted 1: 1000 in 0.05 M potassium phosphate buffer prior to use for functionalization of CNTFETs.

CNTFETs were functionalized with antibodies using a process previously described for the construction of protein-functionalized CNT sensors[18]. Wet-air oxidation was used to create carboxyl groups on the CNT sidewall by heating the nanotubes to 250 $^0$C in a steam atmosphere. EDAC and sNHS were used to activate the carboxyl and link the antibody to the CNT[23]. Figure 1 demonstrates the results of this attachment mechanism.

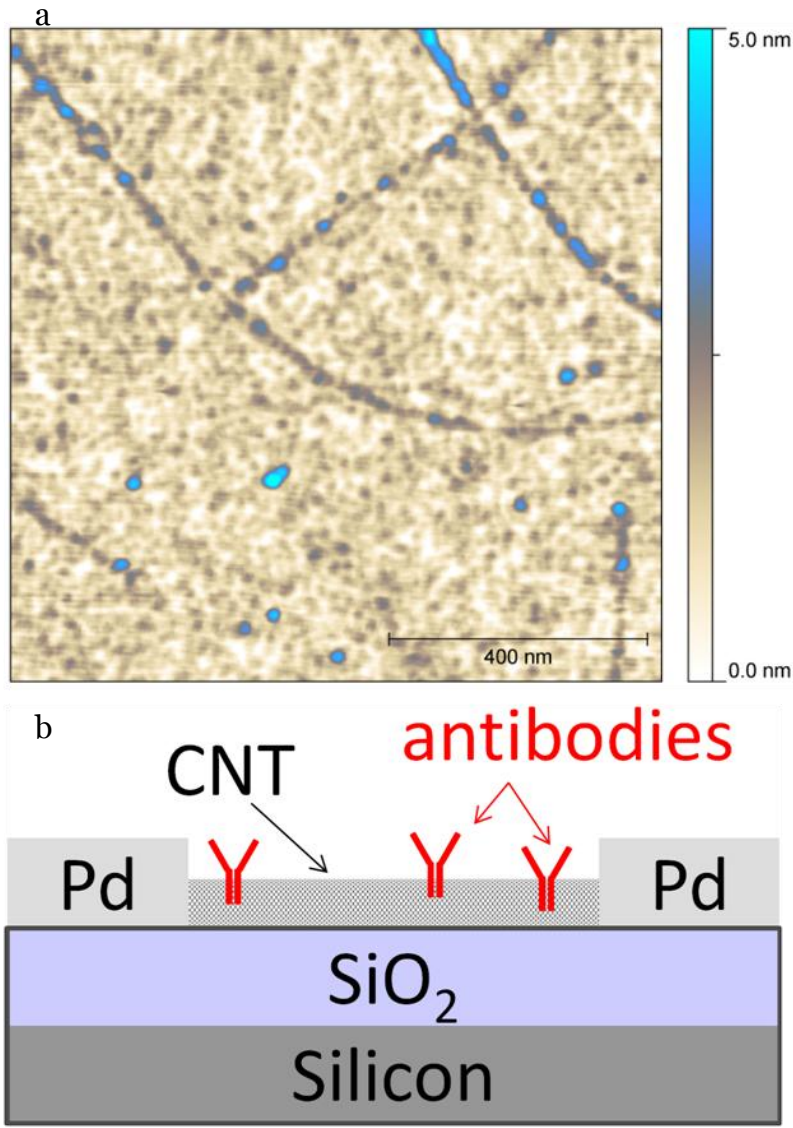

**Figure 1**: Anti-*Salmonella* antibody attachment to CNTs. (a) Image of antibodies and CNTs is taken via Atomic Force Microscopy (AFM). Carbon nanotubes are clearly visible at 1-2 nm higher than the background. The tallest features are anti-*Salmonella* antibodies, roughly 5 nm in height. (b) A diagram of the resulting devices, where a pre-fabricated single nanotube transistor has been functionalized with anti-*Salmonella* antibodies.

Throughout the functionalization process, the electrical properties of CNTFETs were monitored using I-$V_g$ measurements[14, 24]. In the standard measurement protocol, 100 mV was applied across the source and drain electrodes and the DC current through the CNTFET was recorded as the gate voltage was varied from -10 V to 10 V. Figure 2 shows typical I-$V_g$ curves for a CNT FET device after each step of the functionalization process and after *Salmonella* exposure.

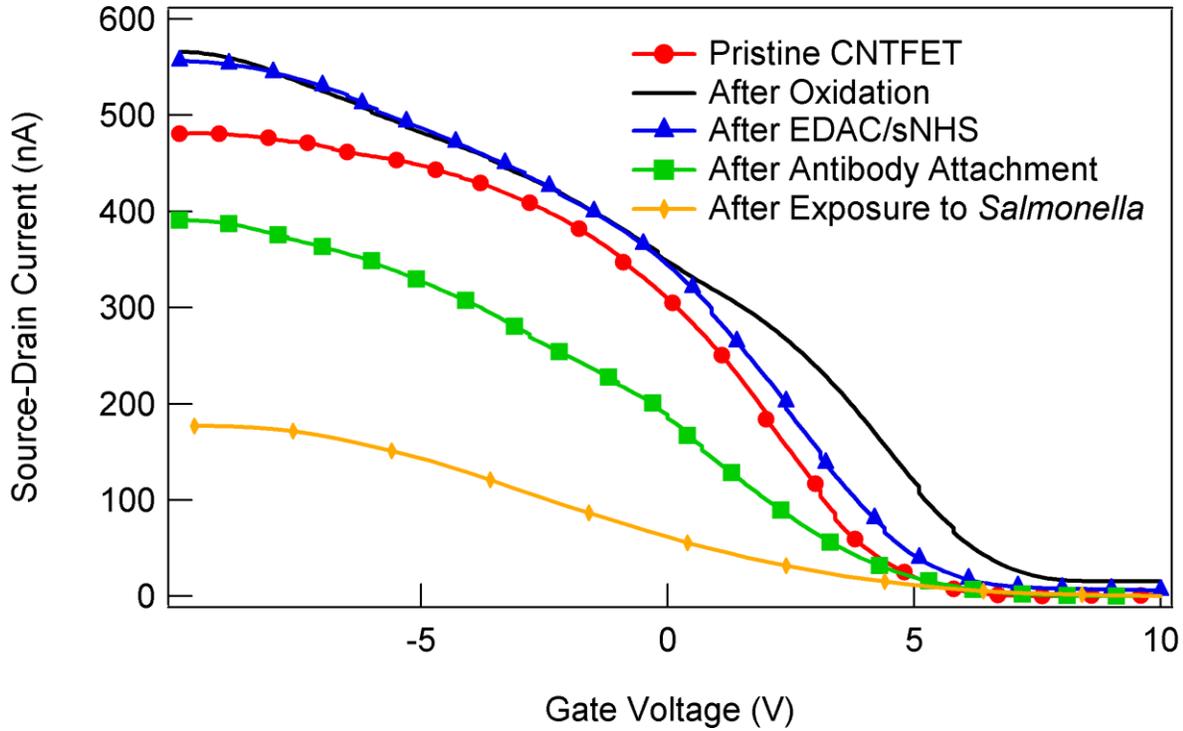

**Figure 2:** I-$V_g$ curves demonstrating *Salmonella* sensitivity. I-$V_g$ curve of a typical un-functionalized CNTFET (red), after wet-air oxidation (black), after EDAC/sNHS treatment (blue), after anti-*Salmonella* antibody treatment (green) and after exposure to $10^7$ cfu/mL *Salmonella* bacteria.

We used two parameters to quantify changes in I-$V_g$ curves: the maximum ON-state current at 100 mV source-drain voltage, and the carrier mobility. The carrier mobility is calculated from the maximum slope of the I-Vg curve and the device geometry using equation 1 with a channel length (*L*) of 2 μm, and calculated backgate capacitance ($C_g$) of 340 fF/cm[25]:

$$\mu = \frac{dI}{dV_g} \frac{L}{C_g} \qquad (1)$$

The changes in I-Vg induced by the functionalization process (Figure 2) were consistent with those from our earlier experiments based on the same attachment chemistry [18]. There is typically a slight increase in the on state current after the steam oxidation process and a positive shift in the threshold voltage of the FET. This shift is consistent with the addition of negatively charged functional sites on the sidewall of the CNT, such as hydroxyl or carboxyl groups in the presence of adsorbed water. Activation of carboxyl groups with EDAC/sNHS leads to a slight decrease in on state current, and a recovery toward the original threshold voltage as some of the charged groups created during oxidation are replaced with an active ester. Addition of the antibody leads to a decrease of ON-state current of 10-50%, due to increased electron scattering from a more complex and crowded chemical environment along the CNTFET conduction channel. The mobility of a CNTFET device generally decreases by a small amount at each of these steps, as additional sources of charge carrier scattering are introduced.

CNTFET devices were exposed to *Salmonella* solutions at various concentrations, prepared as described above, in an incubator at $37^0$C for 45 minutes. It is notable that *Salmonella* was presented in the original growth media with no pre-purification or separation step, in contrast to earlier work [7]. After incubation, CNTFET devices were carefully rinsed with deionized water and electrically characterized by measuring I-Vg curves. Twenty-one devices were successfully functionalized with antibody and tested in *Salmonella* solutions. In order to simplify the experimental interpretation, each CNT device was tested against a single *Salmonella* solution; thus it was possible to test multiple devices at each of six different *Salmonella* concentrations, $10^3 – 10^8$ cfu/ml. Exposure to *Salmonella* at all concentrations tested resulted in a constant dramatic decrease (~ 80%) in on state current, similar to results from prior work on CNT-based *Salmonella* detection suggesting that saturation of the current response occurs at 100-500 cfu/ml[7]. As seen in figure 3, a concentration dependent decrease in mobility is reliably seen, starting at $10^5$ cfu/ml. Thus, in contrast to the ON-sate current, the device mobility is sensitive to bacteria at rather high concentrations.

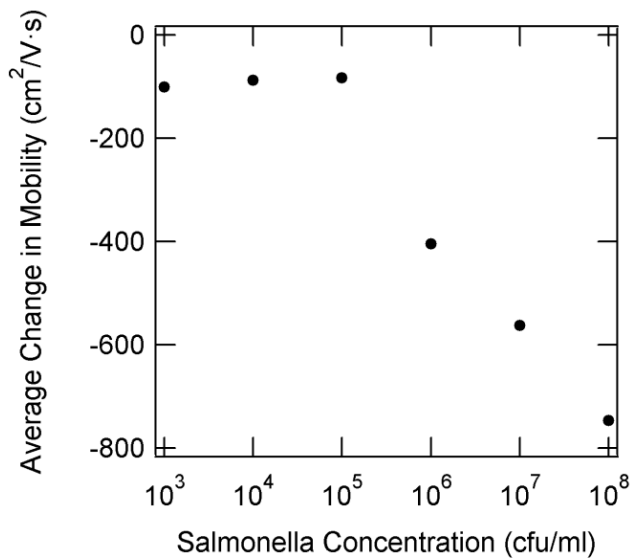

**Figure 3:** Response of CNTFET sensors to varying concentrations of *Salmonella*. Exposure to higher concentration of *Salmonella* is correlated with a larger reduction in carrier mobility.

In control experiments to test the specificity of the sensor, multiple anti-*Salmonella*-functionalized CNTFETs were exposed to three other bacteria: *Escherichiacoli*, *Staphylococcus aureus* and *Streptococcuspyogenes*. When exposed to very high concentrations ($10^8$ cfu/ml) of these three controls, the CNTFETs showed only a small decrease in the ON-state current (ranging from 2-35 %). This observation suggests that this quantity is a direct probe of the bacteria-specific anti-*Salmonella*/*Salmonella* interaction. In contrast, the observed decrease in carrier mobility is very similar for all bacteria tested. Thus, simultaneous measurement of these two parameters could in principal be used to determine both the concentration of *Salmonella* and the concentration of all bacteria in a particular sample of broth.

**Table 1:** Changes in electrical properties of functionalized CNTFETs when exposed to $10^8$ cfu/ml of various bacteria.

| Bacteria | Change in On-state Current (%) | Change in Mobility ($cm^2/V\ s$) |
|---|---|---|
| functionalized with *Salmonella* antibodies | | |
| *Salmonella enterica* serovarTyphimurium | -81 | -747 |
| *Escherichia coli* | -2.5 | -676 |
| *Streptococcus pyogenes* | -35 | -630 |
| *Staphylococcus aureus* | -10 | -804 |

The large decrease in on-state current for all antibody functionalized sensors when exposed to *Salmonella*, coupled with the much lower responses for non-antibody functionalized devices and devices exposed to control bateria indicate that we are seeing a specific, targeted response in the basic electrical properties of CNTFET sensors due to the addition of the *Salmonella* antibody. We believe that using the mobility of such antibody functionalized CNTFET sensors is a unique and useful addition to this field. Here, tracking the mobility leads to additional information in the form of a concentration dependent, but bacteria independent response.

In sum, we demonstrated the use of first-generation carbon nanotube-based immunosensors for detection of *Salmonella* in nutrient broth solutions that mirror the complexity of a real-world, food-chain, application. Concentration-dependent trends in the carrier mobility and reduction in on state current allow for rapid estimation of the *Salmonella* concentration in an unpurified sample. The sensors' ON-state current value shows good specificity for *Salmonella* over other bacteria, and unfunctionalized nanotube devices are not similarly responsive to the presence of non-specific bacteria. The use of a covalent attachment scheme provides additional control over the nanoscale geometry of the sensor and is expected to provide advantages in longevity and performance. The work represents progress towards the eventual application of such nano-enabled sensors to bacteria detection in actual food samples. The use of low-area active device regions will enable implementation of a large array of antibody-functionalized CNTFETs, specific for a variety of food-borne pathogens allowing for facile, simultaneous detection of the most prevalent harmful organisms.

Acknowledgements: Support from Penn's Nano/Bio Interface Center (NSF NSEC DMR08-32802) is gratefully acknowledged, as is use of its facilities. This research was also supported by funds from the NSF-CREST (HRD-0734232) and NSF-HBCU-UP (HRD-0505872) grants to Alabama State University.